\documentclass[a4paper,12pt]{article}
\usepackage{amssymb}
\usepackage{epsfig}

\begin{document}

\title{Gas Density and the ``Volume''~Schmidt Law for Spiral Galaxies}

\author{O.\,V.~Abramova and A.\,V.~Zasov
\thanks{Sternberg Astronomical Institute,
Universitetskii pr. 13, Moscow, 119991 Russia; {oxana@sai.msu.ru},
{zasov@sai.msu.ru}}}

\maketitle

\begin{abstract}
\hspace{0.6cm} The thickness of the equilibrium isothermal gaseous
layers and their volume densities $\rho_{gas}(R)$ in the disc
midplane are calculated for 7 spiral galaxies (including our
Galaxy) in the frame of self-consistent axisymmetric model. Local
velocity dispersions of stellar discs were assumed to be close to
marginal values necessary for the discs to be in a stable
equilibrium state. Under this condition the stellar discs of at
least 5 of 7 galaxies reveal a flaring. Their volume densities
decrease with $R$ faster than $\rho_{gas}$, and, as a result, the
gas dominates by the density at the disc periphery. Comparison of
the azimuthally averaged star formation rate $SFR$ with the gas
density shows that there is no universal Schmidt law $SFR \sim
\rho_{gas}^n$, common to all galaxies. Nevertheless, $SFR$ in
different galaxies reveals better correlation with the volume gas
density than with the column one. Parameter $n$ in the Schmidt law
$SFR \sim \rho_{gas}^n$, formally calculated by the least square
method, lies within 0.8\,--\,2.4 range and it's mean value is
close to 1.5. Values of $n$ calculated for molecular gas only are
characterized by large dispersion, but their mean value is close
to 1. Hence the smaller $\rho_{gas}$ the less is a fraction of gas
actively taking part in the process of star formation.

To be published in Astronomy Reports, 2008.

\end{abstract}

\section{INTRODUCTION}

\hspace{0.6cm}The key question of galaxy evolution is the
dependence of star formation rate $SFR$
 on the gas density and other interstellar medium
parameters, averaged by large enough area or  volume for smoothing
random fluctuations in gas and young stars distributions.

Scmidt~\cite{Schmidt59}  suggested  a simple form of star
formation rate parameterization: $SFR_v\sim\rho_{gas}^n$, usually
called the Schmidt law. From the analysis of gas and young objects
distribution in the solar vicinity, Schmidt~\cite{Schmidt59}
obtained $n\approx 2$. Later a large number of papers were
published aimed to check and to interpret the Schmidt law, based
on the gas and $SFR$ distributions in the discs of  spiral
galaxies. In particular, it was found that both local, azimuthally
averaged $SFR$ values as well as the total star formation rate in
a galaxy correlate (although not too tightly) with the gas content
in its disc (e.g. Madore et al., ~\cite{Madore74}, Kennicutt,
~\cite{Kennicutt89}, Wong\&Blitz, ~\cite{WB02}, Boissier et al.,
~\cite{Bois03}, Shuster et al.,~\cite{Schu06}). Being empirical by
it sence, the Schmidt law and its modifications open a possibility
to calculate evolution models of galaxies, parameterizing the star
formation history. This allows us to understand better  mechanisms
which regulate the rate of star formation.

A situation is entangled by the fact that the dependence
$SFR_v(\rho_{gas})$ (i.e. the local Schmidt law) and, in
particular, the power $n$ can't be found directly from
observations of other galaxies, because in order to estimate the
volume-related values of $SFR_v$ and $\rho_{gas}$ it is necessary
to know the gas layer thickness, which may vary significantly both
along the galaxy radius and from one galaxy to another. Therefore
in practice the Schmidt law is often replaced by the other one,
outwardly alike empirical law $SFR_s\sim\sigma_{gas}^N$ (sometimes
it is called Kennicutt\,--\,Schmidt law), where the compared
values are scaled to unit disc surface area, or by the law
$SFR_t\sim M_{gas}^N$ for the total values $SFR_{t}$ and gas mass
$M_{gas}$ (so-called the global Schmidt law). These dependencies
are more complicated for interpretation, because the compared
parameters are the integrals of heterogeneous functions of
distribution along of the line-of-sight (in the first case) or
allover a whole disc (for the global law). In  general case,
parameters $n$ and $N$ must not coincide (see the discussion of
the question in papers Madore et al., ~\cite{Madore74}, Tutukov,
~\cite{tut06}). Only if at any distance from disc plane $n=1$, the
power $N$ is also equal to unit. In most cases values of $N$ that
obtained for different galaxies lay within the limits of $1<N<2$
(Wong and Blitz, ~\cite{WB02}, Boissier et al.,~\cite{Bois03},
Shuster et al.,~\cite{Schu06}), but for some galaxies they proves
to be more steep. For example, for M~33 $N>$3 (Heyer et al.,
~\cite{HeyerAll04}). For the case of the global Schmidt law when
different galaxies are compared, $N\approx 1.4-1.5$ (Kennicutt,
~\cite{Kennicutt98}, Li et al., ~\cite{li+06}). The scatter  of
$N$ remains large: for galaxies with similar gas masses $SFR_{t}$
may differ by an order of magnitude.

Apparently, it is possible to reproduce the Schmidt law for
different star formation models using some simplifying assumptions
on mechanisms of self-regulation of large-scale star formation, or
considering conversion of neutral gas into dense molecular clouds
and formation of stars (see e.g.,
\cite{tut06,li+06,elm02,krumholz05,Gerristen97}). Note however
that usually  variation of the thickness of gas layer along
galactic radius as well as from one galaxy to another is ignored,
and $\rho_{gas}$ is accepted to be much smaller than  the volume
stellar density, that is not always correct.

In the first part of this paper, the gas volume density in the
plane of an equilibrium disc is calculated as a function of $R$
for several nearby galaxies. In the second part, the relation
between the gas density and star formation rate is analyzed.
Chosen galaxies strongly differs by their properties. They
include: M~33 and M~101 --- multiarmed late-type galaxies (Sc),
the first of them is rather small; interacting galaxy M~51 and
galaxy M~100 which are distinguished by high molecular gas
content; Seyfert galaxy M~106 where the ejections from nucleus and
star formation burst in the inner region are observed; massive
early-type Sab galaxy M~81 which possesses a large bulge and
regular spiral structure, and, finally, our Galaxy. The main
simplifications we use are the following. The gaseous layers in
galaxies are assumed to be axisymmetric and being in hydrostatic
equilibrium. The pressure of the gas is determined by its
turbulent motion: $P_{gas}=\rho_{gas}\,C_z^2$, where $C_z$ is
one-dimensional velocity dispersion, which is assumed to be
constant (although different for atomic and molecular gas). These
simplifying suggestions are definitely too tough for  regions
enveloped by the intensive star formation, for the inner discs
deep inside dense bulges or in the neighborhood of active nucleus,
and also for the far periphery of discs. Note however that gas
velocity dispersion, although may slowly vary with the distance
from the galaxy center, remains high enough even at large
distances (see the discussion in paper Dib et al., ~\cite{dib06}).
Magnetic field pressure gradient and thermal gas pressure play
significantly lesser role in the formation of gas layer thickness,
at least for the case of our Galaxy (see  discussion of this
question in Cox review,~\cite{Cox05}). It is essential that within
the limits of approximations mentioned above the observed
distribution of atomic (HI) and molecular (H$_2$) hydrogen disc
thicknesses along the Galaxy's radius can be sufficiently
explained (Narayan and Jog,~\cite{NJ02}).

\section{Stellar and Gaseous Discs Thicknesses}

\hspace{0.6cm}The gaseous disc thickness depends on the turbulent
gas velocity and  gradient of gravitational potential along the
$z$-coordinate, perpendicular to disc plane. The latter is a sum
of gravitational potentials of all galaxy components. Therefore
for the gas density estimation one have to know not only
individual component masses, but also the vertical profile of
stellar disc density. Both stellar and gaseous disc scale heights
are much less than their radial scales. It allows to ignore the
influence of the radial heterogeneity of discs on the vertical
gradient of potential.

Our approach to the estimation of the vertical gaseous and stellar
density distributions  is similar to that described by Narayan and
Jog,~\cite{NJ02}. Hydrostatic equilibrium equation can be written
as
\begin{equation}\label{10000}
\frac{<(C_z)_i^2>}{\rho_i}\,\frac{d\rho_i}{dz}\, =\,(K_z)_*+
(K_z)_{\rm HI}+(K_z)_{\rm H2}+(K_z)_{\rm DM}.
\end{equation}
Here $\rho$ is the volume density, index $i$ (asterisk, $\rm
{HI\,,H_2}$) corresponds to the stellar, atomic, or molecular
discs, and $(C_z)_i$ is vertical dispersion of velocities.

Under  assumptions  already mentioned, the stellar, atomic, and
molecular volume density distributions are described by the
equation that follows from condition of the hydrostatic
equilibrium (\ref{10000})\, (equation (3) in paper~\cite{NJ02}):
\begin{equation}\label{100}
\frac{{\rm d}^2\rho_i}{{\rm d}z^2}\,=\frac{\rho_i}{<(C_z)_i^2>}\,
\left[-4\pi G\,(\rho_*+\rho_{\rm HI}+\rho_{\rm H_2})+\frac{{\rm
d}(K_z)_{\rm DM}}{{\rm d}z}\right]+
\frac{1}{\rho_i}\,\left(\frac{{\rm d}\rho_i}{{\rm d}z}\right)^2\,.
\end{equation}
Here ${\rm d}(K_z)_{DM}/{\rm d}z=\partial^2\psi_{\rm DM }/\partial
z^2$ describes the input of dark halo, where
$$
\frac{\partial^2\psi_{\rm DM }}{\partial
z^2}=\frac{v_{rot}^2\,R_C}{(R^2+z^2)^\frac32}\arctan\left(\frac{\sqrt{R^2+z^2}}{R_C}\right)\,
\left[1-\frac{3z^2}{R^2+z^2}\right]+
$$
$$
+\frac{z^2R_C^2v_{rot}^2}{(R^2+z^2)^2(R_C^2+R^2+z^2)}+
\frac{v_{rot}^2}{(R^2+z^2)}\left[\frac{2z^2}{(R^2+z^2)}-1\right]~-
$$
is the $z$-component of the  second derivative of the halo
potential in cylindrical coordinates~(\cite{NJ02}). The density
distribution in the halo is accepted to be quasi-isothermal:
\begin{equation}\label{105}
\rho_{\rm DM}(R)=\frac{v_{rot}^2}{4\pi
G}\,\frac{1}{\left(R_C^2+R^2\right)}~.
\end{equation}
Here $R$ is the distance from the center of  galaxy, and $v_{rot}$
and $R_C$ are the circular velocity and nucleus radius of the
halo, respectively.

Equations~(\ref{100}), applied to each galaxy component, can be
reduced to the first order equations. Their self-consistent
solutions are found by fourth-order Runge\,--\,Kutta method. Two
boundary conditions required for the mid-plane, $z=0$ are:
\begin{equation}\label{103}
\rho_i=(\rho_0)_i~,~\frac{{\rm d}\rho_i}{{\rm d}z}=0~.
\end{equation}
Values of the central densities $(\rho_0)_i$ can be determined
from the following evident condition
\begin{equation}\label{104}
2\,\int_0^\infty\rho_i(R,z){\rm d}z=\sigma_i(R)~,
\end{equation}
(here $\sigma_i(R)$ is the observed radial distribution of
surface, or column, densities of corresponding components) and are
found by the bissection algorithm. Initially,
equations~(\ref{100}) are solved for stellar disc assuming
$\rho_{\rm HI,H_2}=0$. After that, using the obtained solution,
equation for HI is solved with $\rho_{\rm H_2}=0$. Finally,
equation for molecular hydrogen is solved using values obtained
earlier for HI and stars. This procedure is iterated using
$\rho_i$ obtained in previous iterations, until the solution
converges. Thickness of discs is calculated as doubled
half-width-half-maximum (HWHM) value of corresponding density
distribution.

Indeed, to find the gas density by the method described above, it
is necessary  to calculate masses of  each component of a galaxy
as well as to find stellar disc velocity dispersion, which varies
significantly with $R$ even for a single galaxy unlike the gas
velocity dispersion.

For all galaxies from our sample, except our Galaxy, stellar disc
surface densities were estimated from the available  data on the
radial surface brightness distribution in disc and disc's integral
color index corrected for galactic absorption and inclination
(based on the HYEPERLEDA~\cite{Leda} database or original
photometry  taken from the literature). We ignored the central
parts of galaxies where the bulge dominates and/or observed
circular rotation curve is uncertain. The largest values of $R$ we
considered were limited by extention of measured rotation curve or
(M~101, M~106)
 gas surface density distribution. To convert
the surface brightness to stellar disc surface density, we used
the mass-to-luminosity ratios corresponding to color index of
discs (Table A3 in Bell \&de Jong  ~\cite{BdeJ01}). These ratios
were used as the first approximation to decompose the rotation
curve in the frames of three-component model of galaxies which
consist of the bulge, exponential disc, and quasi-isothermal halo.

The accepted distances to the galaxies, disc inclinations, and the
annuli, for which we compared  $SFR$ and gas density are presented
in Table~\ref{tab1}.
\begin{table}[h]
\caption{Properties of the  galaxies. \emph{Columns}: (1)
--- galaxy, (2) --- distance, (3) --- inclination, (4) ---
major axis diameter at the B\,=\,25\,$^m$/$\square''$ isophote,
(5) --- radial interval, for which parameter $n$ in Schmidt law is
derived.}\label{tab1}
\begin{center}
\begin{tabular}{ccccc}
&&&&\\
Galaxy&$\cal
D$&$\textit{i}$,\,$^\circ$&$D_{25}/2$&$\Delta R_{Schmidt}$\\
&Mpc&&&kpc\\
(1)&(2)\,&(3)\,&(4)\,&(5)\\
&&&&\\
\hline
&&&&\\
M~33&0.70&55&$35.4'$&~1,0~--~6,1~\\
&&&&\\
M~51&8.4&20&$5.6'$&~2,0~--~8,3~\\
&&&&\\
M~81&3.63&59&$13.45'$&~4,0~--~11,6~\\
&&&&\\
M~100&17.00&27&$3.7'$&~2,0~--~16,3~\\
&&&&\\
M~101&7.48&21&$14.4'$&~2,0~--~10,9~\\
&&&&\\
M~106&7.98&63&$9.3'$&~2,1~--~9,5~\\
&&&&\\
Galaxy&---&---&---&~3,0~--~13,5~\\
&&&&\\
\hline
\end{tabular}
\end{center}
\end{table}
Table~\ref{tab2} demonstrates the color index (which was used for
the mass-to-luminosity ratio), radial brightness scale lenght and
corresponding references, references to the sources of used
rotation curves.
\begin{table}[h]
\caption{Input data related to galaxy discs. \emph{Columns}: (1)
--- galaxy, (2) --- Color index, used in the disc mass calculations and
reference to original source, (3) -- radial scale length and
reference to original source, (4) --- reference to original source
of rotation curve, (5) --- reference to the $HI$ distribution
source, (6) --- reference to the $H_2$ distribution
source.}\label{tab2}
\begin{center}
\begin{tabular}{lccccc}
Galaxy&\footnotesize{Color}&\footnotesize{Brightness}&\footnotesize{Rotation}&
\footnotesize{Atomic}&\footnotesize{Molecular}\\
&\footnotesize{index}&\footnotesize{scale}&\footnotesize{curves}&
\footnotesize{hydrogen}&\footnotesize{hydrogen}\\
(1)&(2)&(3)&(4)&(5)&(6)\\
&&&&&\\
\hline
&&&&&\\
M~33&$(V-I)$~\cite{Lauer}&$5.8'$~\cite{RegVog}&\cite{Corb}&\cite{Corb}&\cite{Corb}\\
&&&&&\\
M~51&$(B - V)$~\cite{Leda}&$87,4''$~\cite{Baggett}&\cite{STHTTKT99}&\cite{main}&\cite{main}\\
&&&&&\\
M~81&$(B-V)$~\cite{Leda}&$158''$~\cite{Baggett}&\cite{STHTTKT99}&\cite{Rots}&\cite{main}\\
&&&&&\\
M~100&$(V-I)$~\cite{Beckman}&$48.5''$~\cite{de Jong}&\cite{STHTTKT99}&\cite{main}&\cite{main}\\
&&&&&\\
M~101&$(B-V)$~\cite{Leda}&$128''$~\cite{Knapen}&\cite{STHTTKT99}&\cite{WB02}&\cite{WB02}\\
&&&&&\\
M~106&$(B-V)$~\cite{Leda}&$163''\,\,^*$&\cite{vanAlb80}&\cite{main}&\cite{main}\\
&&&&&\\
\hline
&&&&&\\
\end{tabular}
\end{center}
$^*$\,\footnotesize{In~\cite{Flores} the disc scale in the optical
range $b_{opt}$ is given as 5,22$\,kpc$ or $163''$ for the
accepted distance to M~106. From the brightness
distribution~\cite{HeraudSimien,Sanchezetal}, it follows that
radial disc scales in different wavebands for this galaxy are
practically similar, so in this paper we used $r_0=163''$.}
\end{table}
The resulting model parameters are shown in Table~\ref{tab3}. They
include the disc and halo parameters, and the star formation
rates, integrated over azimuth within the distance intervals
$\Delta R$.
\begin{table}[h]
\caption{Model parameters of galaxies discs and halos and star
formation rates. \emph{Columns}: (1) --- galaxies, (2) --- linear
radial scale length, (3) central surface density of a disc in a
model, (4) --- asymptotical velocity, (5) --- halo nucleus radius,
(6) --- star formation rates $SFR_{int}$  (7) --- radial
intervals, for which $SFR_{int}$ is related.}\label{tab3}
\begin{center}
\begin{tabular}{c|cc|cc|cc}
&&&&&&\\
Galaxy&\multicolumn{2}{c|}{Disc}&\multicolumn{2}{c|}{Halo}&$SFR_{int}$&$\Delta R$\\
&\multicolumn{2}{c|}{parameters}&\multicolumn{2}{c|}{parameters}&$\cal M_\odot$/year&kpc\\
&&&&&&\\
&\footnotesize{Scale length, kpc}&\footnotesize{$\sigma_0,~\cal M_\odot$/pc$^2$}&$v_\infty$\,,km/sec&$R_c$,\,kpc&&\\
(1)&(2)\,&(3)\,&(4)\,&(5)\,&(6)\,&(7)\,\\
&&&&&&\\
\hline
&&&&&&\\
M~33&1,2&657,5&109,8&1,41&0,4&0~--~6,5\\
&&&&&&\\
M~51&3,6&1238&120&3,25&5,9&0,5~--~17,4\\
&&&&&&\\
M~81&2,8&1710&88&4,6&1,3&0,1~--~11,6\\
&&&&&&\\
M~100&4,0&1175&295,7&5,3&6,5&0~--~21,8\\
&&&&&&\\
M~101&4,6&629&236&5,2&5,3&0,7~--~27,9\\
&&&&&&\\
M~106&6,3&934&157&8&3,3&0,7~--~20,2\\
&&&&&&\\
Galaxy&3,2&641&220&5&3,7&3~--~13,5\\
&&&&&&\\
\hline
\end{tabular}
\end{center}
\end{table}
For our Galaxy, the halo and stellar disc parameters and stellar
disc HI and H$_2$ dispersions are assumed the same as in the
Narayan and Jog's paper~\cite{NJ02}, where the velocity dispersion
supposed to be constant for the gas, and exponentially decreasing
with $R$ for stars..

To estimate stellar disc thickness, two independent methods are
employed in this paper. In the first method, the stellar velocity
dispersion  $C_z$ is considered to be proportional to the minimum
value of the radial velocity dispersion $C_r$, which provides
gravitational disc with stability to
 perturbations in its plane. Direct measurements of the
velocity dispersion of old stellar discs support the suggestion
that local values of the velocity dispersion in spiral galaxies
are usually close to the minimal ones necessary to provide the
dynamical stability to the disc (Bottema,~\cite{Bottema93}, Zasov
et al., ~\cite{ZasAll04}). The expression for stellar velocity
dispersion along the $z$-coordinate can be written as:
\begin{equation}\label{108}
C_{z*}(R)=K\cdot3,36\cdot\frac{G\cdot\sigma(R)\cdot
Q(R)}{\kappa(R)},
\end{equation}
where $K\approx$ 0,4\,--\,0,7 is the velocity dispersion ratio
$C_z/C_r$, determined by the stability disc condition against the
bending modes, $\sigma(R)$ designates the surface disc density,
$\kappa(R)=\sqrt{2\,\frac{v(R)}{R}\left(\frac{v(R)}{R}+\frac{dv}{dR}\right)}$
--- epicyclic frequency for the rotation curve $v_{rot}(R)$. Strictly speaking, stability parameter Toomre
$Q(R)$ is equal to unit only for the idealized case of thin
homogeneous disc exposed to radial perturbations. N-body modeling
of marginally stable disc with various input data (see
e.g.~\cite{HZT03} and references in this paper) demonstrates that
the $Q$- parameter varies along the radius in the range of
1.2\,--\,3 for a large part of disc within several radial scale
lengths. In this paper we used the approximation formula $Q(R)$
from~\cite{HZT03}:
\begin{equation}\label{106}
Q^*_T=A_0+A_1\,\left(\frac{R}{R_0}\right)+A_2\,\,{\left(\frac{R}{R_0}\right)}^2\,,
\textrm{where}
\end{equation}
$A_0=1,46$, $A_1=-0,19$, $A_2=0,134$; where $R_0$ is radial disc
scalelength.

The epicyclic frequency is calculated for M~33 and M~100 using the
observed rotation curve. After that, the resulting radial
dependence of disc thickness was smoothed . For the galaxies M~51,
M~81, M~101 and M~106, where the irregularities of rotation curves
$v_{rot}(R)$ may lead to considerable errors in $\kappa$ , we use
the initially``smoothed'' rotation curves, calculated for
three-component model galaxies (best fit models). For all galaxies
(except the Galaxy) the values of $C_{z*}(R)$, calculated from the
equations above, are used to estimate the half-thickness of
stellar disc $h_*(R)$ and volume densities of stellar and gaseous
discs in the galactic plane. Parameter $K$ in equation~(\ref{108})
is assumed to be 0.5.

\begin{figure}[h]
\epsfig{file=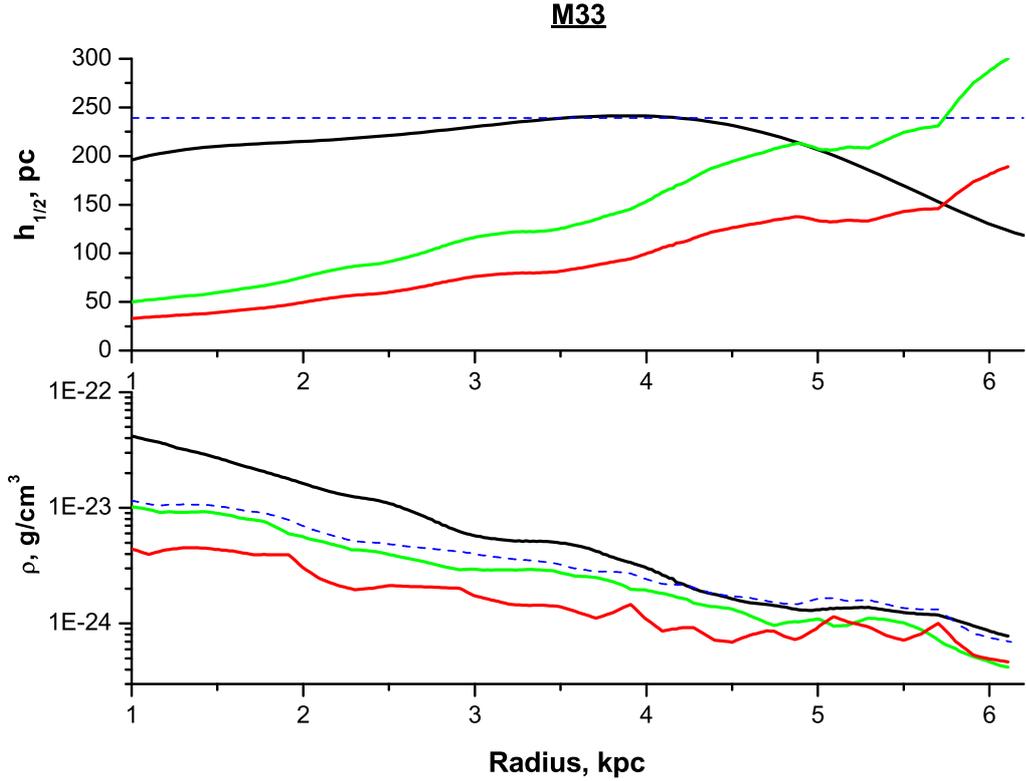,height=12cm,width=15cm} \caption{Model
radial dependencies for M~33. Above: a half thicknesses of the
stellar disc and gas layers, below --- their volume densities in
the midplane. Black thick continuous lines correspond to stellar
disc, green lines --- to HI layer, red lines --- to H$_2$ layer;
blue dash thin lines are fit the model of constant stellar disc
thickness: a half thicknesses of the stellar disc in the figure
above and the total gas density  in the figure below.}
\label{fig1}
\end{figure}

The results of calculation of the half-thickness (defined as HWHM)
of stellar disc $h_*(R)$, and HI and H$_2$ layers are shown in
figures\,\ref{fig1}a\,--\,\ref{fig7}a (upper pannel). Radial
dependence $h_*(R)$ for our Galaxy practically coincides with
Narayan\&Jog results~\cite{NJ02} because the same equations and
input data are used (though mathematical method of solution of the
equations is different). Note, however, that $h_*(R)$ for the
Galaxy in our paper is extended farther from the center. The
results for molecular gas are also noticeably different from those
derived by Narayan and Jog because the input data were adopted
from another source.

\begin{figure}[h]
\epsfig{file=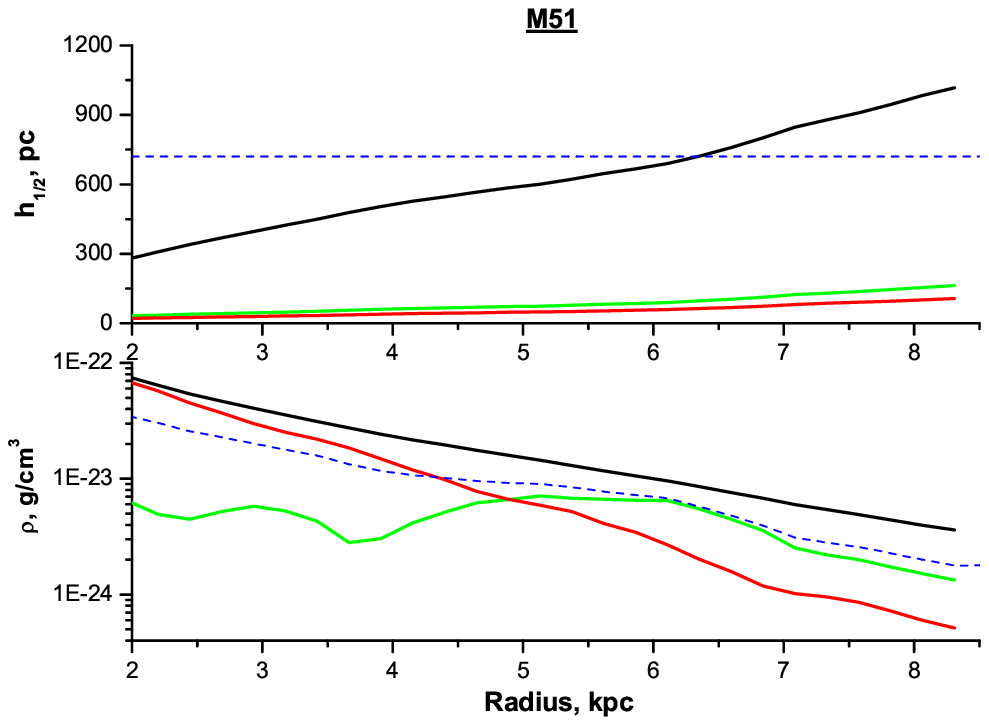,height=12cm,width=15cm} \caption{Model
radial dependences for M~51. Designations are the same as on the
Figure~1.} \label{fig2}
\end{figure}

\begin{figure}[h]
\epsfig{file=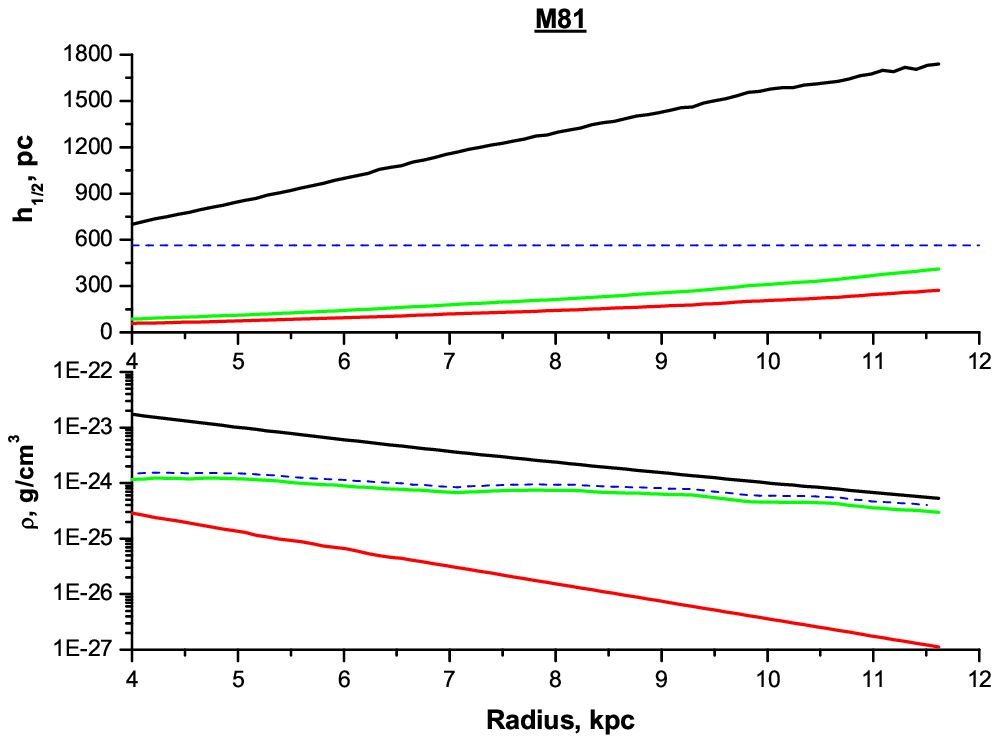,height=12cm,width=15cm} \caption{Model
radial dependences for M~81. Designations are the same as on the
Figure~1.} \label{fig3}
\end{figure}

\begin{figure}[h]
\epsfig{file=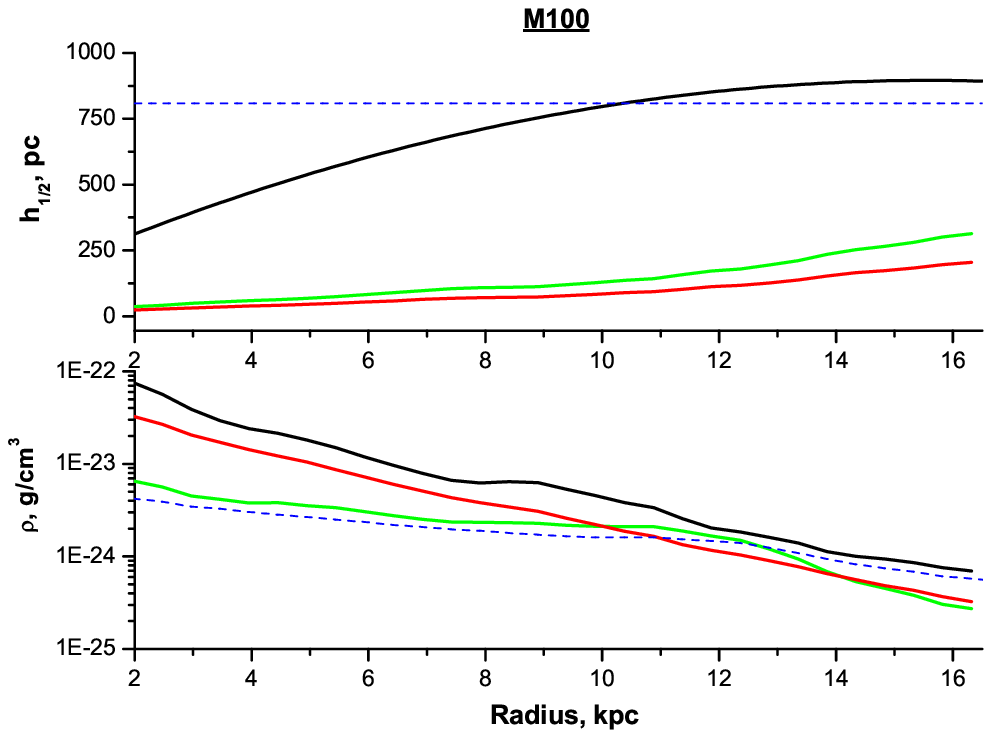,height=12cm,width=15cm} \caption{Model
radial dependences for M~100. Designations are the same as on the
Figure~1.} \label{fig4}
\end{figure}

\begin{figure}[h]
\epsfig{file=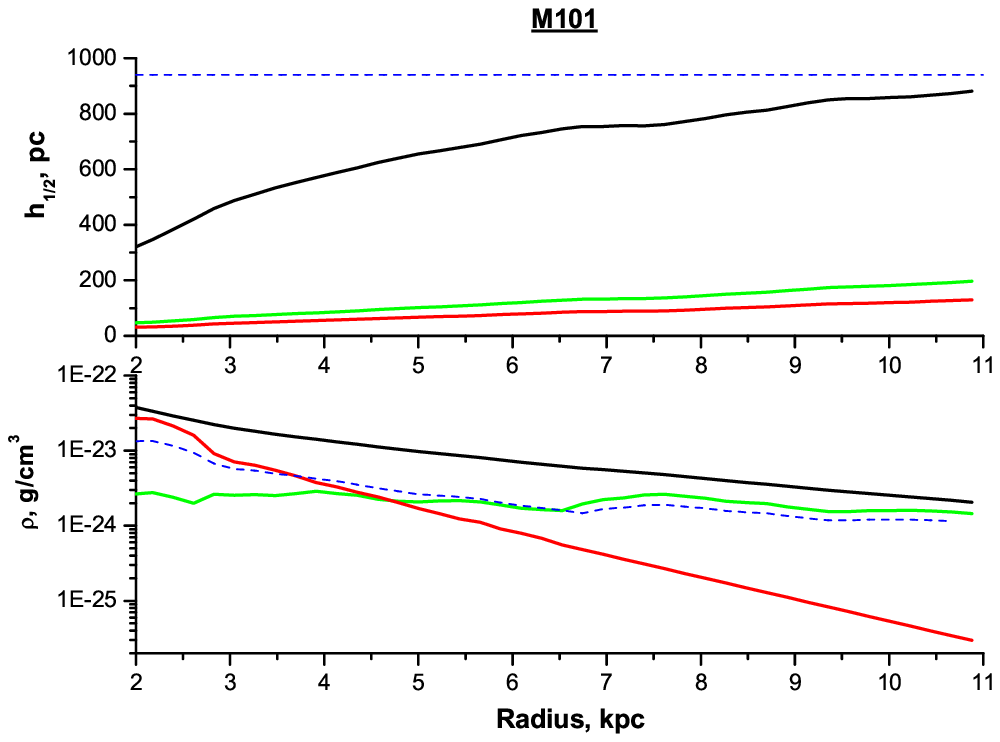,height=12cm,width=15cm} \caption{Model
radial dependences for M~101. Designations are the same as on the
Figure~1.} \label{fig5}
\end{figure}

\begin{figure}[h]
\epsfig{file=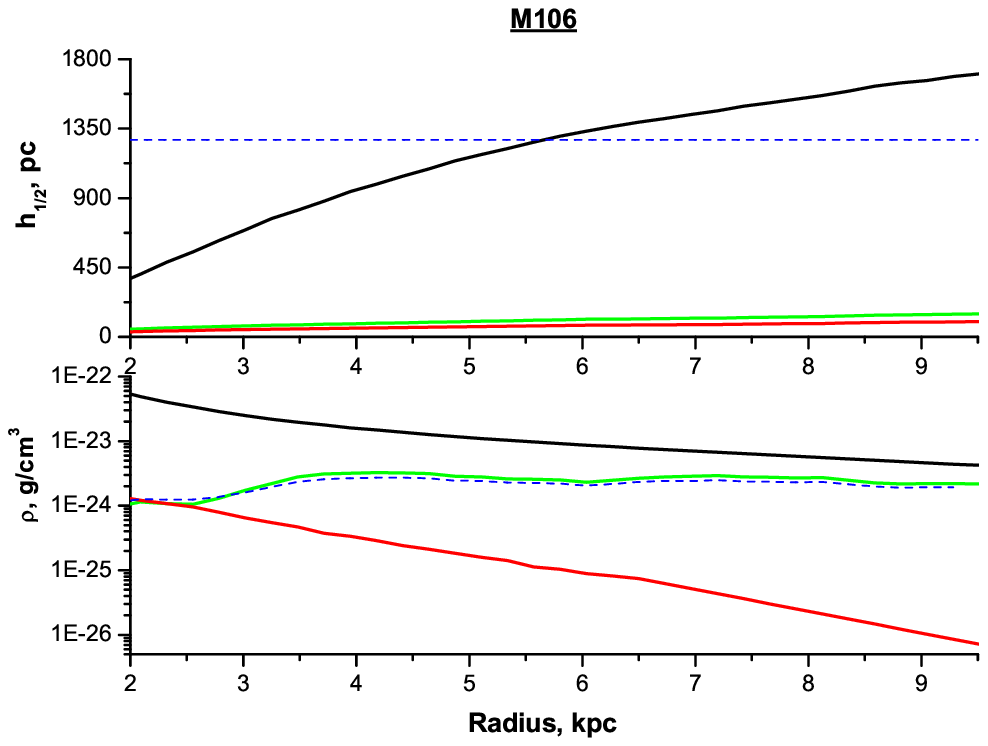,height=12cm,width=15cm} \caption{Model
radial dependences for M~106. Designations are the same as on the
Figure~1.} \label{fig6}
\end{figure}

\begin{figure}[h]
\epsfig{file=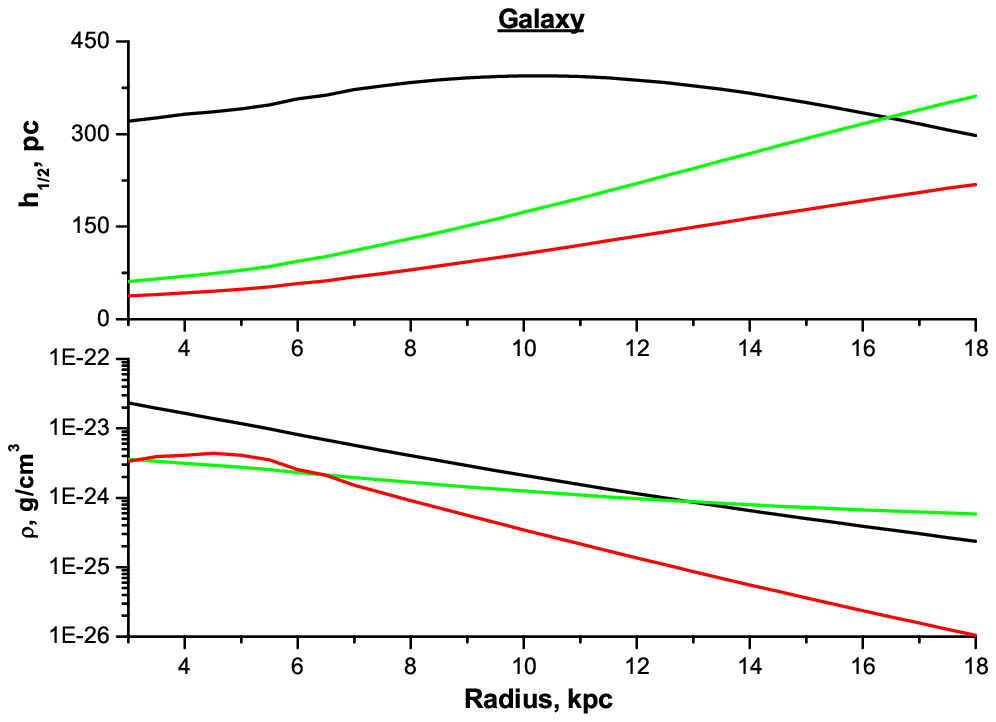,height=12cm,width=15cm} \caption{Model
radial dependences for our Galaxy. Designations are the same as on
the Figure~1 (except of the absence of thin dashed
lines).}\label{fig7}
\end{figure}

Our results show that only for our Galaxy and maybe  M~33 it can
be concluded that the stellar disc thickness weakly changes with
galactocentric distance within a large range of $R$. In M~33, disc
thickness at the periphery falls down. However, this result is not
reliable and needs to be confirmed, because for this galaxy it may
be caused by the smoothing of rather irregular curve. For other
our galaxies the distribution of $h_*(R)$ is determined more
reliably and disc's thicknesses tends to grow with $R$ as it can
be seen from figures~\ref{fig2}\,--\,\ref{fig7} ( top pannels).

The question whether the stellar discs thickness in real galaxies
changes with $R$ seems to be more simply resolved for edge-on
galaxies. Nevertheless, the available estimates obtained from
photometric data are rather discrepant. There are arguments both
for the thickness constancy (see for example van der
Kruit,~\cite{vdKruit88}), and for the disc flaring --- especially
for early-type disc galaxies, where the disc thickness increases
approximately  1.5 times within one radial brightness scalelength
(de Grijs and Peletier, ~\cite{deGrPel99}). The latter conclusion
is in a good agreement with our calculations, at least for
Sab\,--\,Sb galaxies. Our estimations are based on an assumption
of marginal disc stability. Hence, if the discs actually keep
their thicknesses constant, being nevertheless in a dynamical
stable condition, it would mean that our model underestimates the
discs thickness at least in their inner regions, and , as a
sequence, the discs are dynamically overheated there.

In the second method of calculation of the gas densities, we
propose that the half thicknesses of the isothermal stellar disc
$h_*$ is independent of $R$ and is equal to 0.2\,$R_{0}$, where
$R_{0}$ is the photometric disc scalelength. Note that the change
of the accepted value of $h_*/R_0$ influences the resulting gas
density, but weakly affects the shape of the radial density
dependencies in general.

In the model of fixed stellar disc thickness a calculating
procedure should be different because  the solution may not be
self-consistent in this case. If the input of  gas into the
general surface disc density is small, the vertical density
distribution in the isothermal stellar disc can be described by
the well-known equation:
\begin{equation}\label{112}
\rho_*=\rho_{0*}\left(sech\left(\frac{|z|}{z_0}\right)\right)^2\,
\end{equation}
where
\begin{equation}\label{113}
z_0~=~\frac{C_{z*}^2}{\pi G\sigma_{tot}}\,,
\end{equation}
and $\sigma_{tot}$ --- the total surface disc density at given
$R$.

Gravitational field of the stellar disc creates the vertical
acceleration
\begin{equation}\label{114}
K_z= -k^2\cdot th (|z|/z_0), \,
\end{equation}
where $k^2=4\pi G\rho_{0*}z_0$. Gas density equilibrium
distribution (HI or H$_2$) in such a field is described by the equation
~\cite{celnik79}:
\begin{equation}\label{115}
\rho_i(z)
=\rho_{0i}\,\left(sech\left(\frac{|z|}{z_0}\right)\right)^\alpha,
\end{equation}
where $\alpha=k^2z_0/C_{zi}^2$.

Using (\ref{113}) and taking into account that the stellar disc
surface density is $\sigma_* = 2\rho_{0*}z_0$, we get:
\begin{equation}\label{116}
\alpha =2\left(\frac{C_{z*}}{C_{zi}}\right)^2=\frac{2\pi G
\sigma_*z_0}{C_{zi}^2}.
\end{equation}
Joint solution of equations (\ref{104}, \ref {115}, \ref{116})
enables us to estimate the gas density $\rho_{0i}(R)$ in the disc
plane for a given scaleheight $z_0$. For the case of our Galaxy,
the radial gas distribution and velocity dispersions in HI, H$_2$
and stellar discs are taken the same as in Narayan and Jog's
paper~\cite{NJ02}: $C_z({\rm HI})=8$\,km/sec, $C_z({\rm
H_2)}=5$\,km/sec, and for stars an exponential function of $R$ is
accepted following \cite{LewFrem89}):
\begin{equation}\label{107}
C_{z*}(R)\,\textrm{[km/sec]}\approx 50 \cdot
\exp\left\{-\frac{R\,\textrm{[kpc]}}{8,74}\right\}~.
\end{equation}
The velocity dispersion $C_{z*} (R)$ for $R>$3 kpc is defined to
be close to the values obtained in N-body simulations of
marginally stable discs (see Figure~8 in Khoperskov and Tyurina
paper~\cite{Kh03}). For six remaining galaxies we assumed the gas
velocity dispersions along the $z$-axis to be $C_z({\rm
HI})=9$\,km/sec for $HI$ and $C_z({\rm H_2})=6$\,km/sec for $H_2$.

Results of calculation of the volume density in stellar discs, and
 corresponding values of densities in the midplane are
illustrated in figures\,\ref{fig1}\,--\,\ref{fig7} (lower
pannels). Dashed line shows distribution of the total gas density
$\rho_{gas}$ in the model with fixed stellar disc thickness. Note
that both approaches give qualitatively similar results, although
the volume gas density decreases slower along the radius in the
discs with constant thickness.

It is obvious that the gas volume density falls along the radius
faster than the surface density because of gas layer flaring at
large $R$. The density of stellar discs decreases even faster than
gaseous, so these densities become nearly equal at the discs
periphery. For example, in M~33 the gas density prevails over the
stellar one starting with $R\approx$~5\,kpc. In our Galaxy it is
observed satrting with $R\approx$~12\,kpc . Nevertheless, the
stellar disc remains dominated by the surface density over gaseous
one even at this distance  because of the larger half thickness.

\section{Star formation and the Schmidt law}

\hspace{0.6cm}The star formation rate $SFR_s$ per unit of disc
area, and its variation along the radius of spiral galaxies, as
well as the relationship between $SFR_s$ and other galactic
parameters, were examined by many authors (see Introduction for
some references). In this paper  we take radial distributions
$SFR_s(R)$ obtained from the combined UV and far IR brightness
data following the papers by Boissier et al.~\cite{main},
Hirashita et al.\cite{HBI03}, Buat et al.\cite{BBGB02}. We use the
data  from Boissier et al.~\cite{main} where UV absorption
profiles smoothed over $\sim 100''$ are presented. This approach
supposes that all the energy absorbed in UV range reradiates in
FIR, so that brightness ratio UV/FIR is the quantitative criterion
of this absorption. The resulting values of UV brightness
corrected for the absorption weakly depend upon the optical
properties of dust and the spatial distribution of dust and
luminous material, i.e. they are stable against the model
choice~\cite{Iglesias04}. Following~\cite{HBI03,BBGB02}, star
formation rate may be written as:
\begin{equation}\label{5}
SFR=\frac{SFR(UV)}{1-\varepsilon}\,.
\end{equation}
Here $1-\varepsilon$ is a fraction of not absorped UV quanta, that
is estimated by ratio FIR/UV, and $SFR(UV)=C_{2000}\cdot L_{2000}$
is star formation rate, calculated by UV radiation intensity,
$C_{2000}$ is a coefficient derived in stellar population
modeling, and
$L_{2000}\,\left[\frac{\textrm{erg}}{\textrm{sec}\cdot\AA\cdot{\textrm{pc}}^2}\right]$
designates the monochromatic luminosity at 2000$\AA$. For stellar
population model with solar abundance and
 Salpeter initial mass function (IMF) of stars in the interval
0,1\,--\,100\,$M_\odot$ coefficient $C_{2000}$ is taken as
$2,03\cdot10^{-40}\left[\frac{{M}_\odot}{\textrm{yr}\cdot\textrm{erg}/
(\textrm{sec}\cdot\AA)}\right]$~\cite{BBGB02}. Note that the
estimates of $SFR_s$, obtained by the same way, we used in our
previous paper~\cite{ZasAbr06}.

A different way is needed to estimate the star formation rate for
our Galaxy. We tentatively admit that the $SFR$ changes along $R$
parallel to the azimuthally averaged FIR brightness distribution
provided by the radiation of several hundred star forming regions
studied by Bronfman et al.~\cite{Bronfetal00}. The radial
distribution curve (see Figure~10 in that paper) was calibrated in
such a way that for the solar vicinity  $SFR_s$ equals to $4\cdot
10^{-9}$~ M$_\odot$/(yr$\cdot$pc$^2$). Being integrated all over
the galactic disc (within the limits considering in this paper),
it gives the total $SFR_t \approx 3.6~$M$_\odot$/yr, which is in a
good agreement with the available integral estimations of $SFR_t$
\cite{Rana91}.

\begin{figure}[h]
\epsfig{file=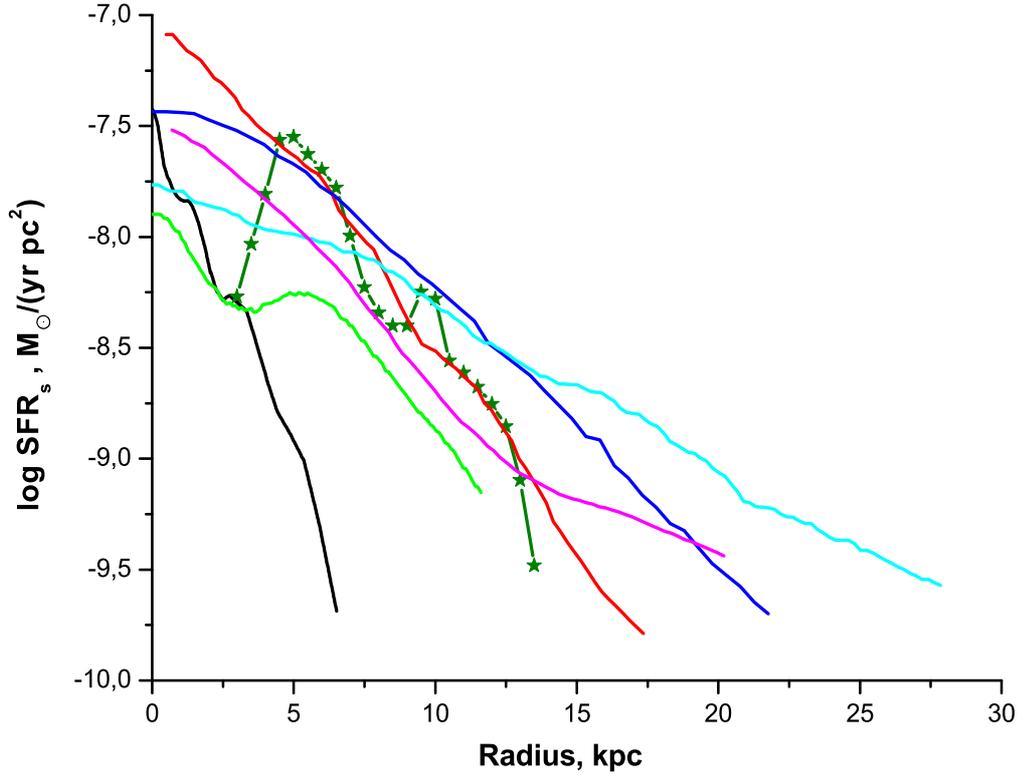,height=12cm,width=15cm} \caption{Radial
dependences of star formation rates $SFR_s$ for galaxies.
Continuous black line --- M~33, red --- M~51, green --- M~81, blue
--- M~100, cyan --- M~101, magneta --- M~106, continuous olive line with stars --- Galaxy.}\label{fig8}
\end{figure}

\begin{figure}[h]
\epsfig{file=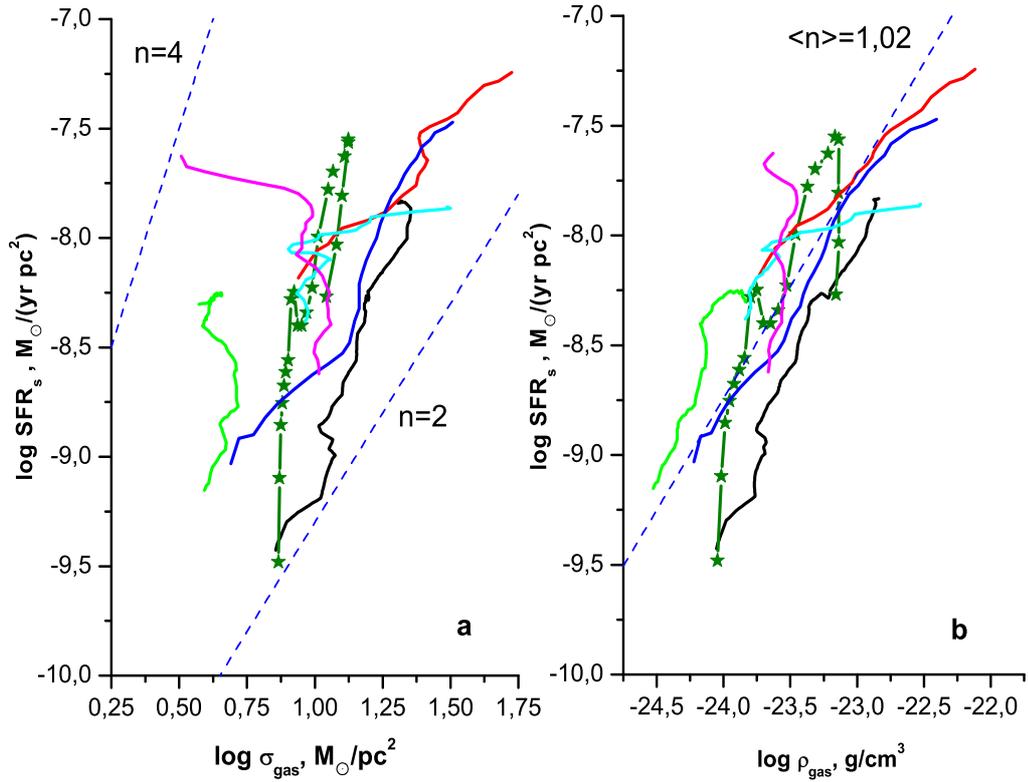,height=12cm,width=15cm}
\caption{Dependence of star formation rates $SFR_s$ on the surface
(left) and volume (right) gas densities. Designations are the same
as on the Figure~8.} \label{fig9}
\end{figure}

Radial dependencies $SFR_s(R)$ for our set of galaxies are shown
in Figure~\ref{fig8}.
In Figure~\ref{fig9}, the star formation rate compares with
surface gas density (Kennicutt\,--\,Schmidt relationship). For the
combined data taken for all galaxies, the relationship really
exists, although the individual curves form a broad bundle.
Similar behavior of graphs can be found in other papers (see e.g.
Figure~10 in paper Boissier et al.~\cite{Bois03}). Note that for
M~81 the dependence is practically absent, and for M~106 it is
ambiguous: in the inner part of the galaxy $SFR_s$ decreases with
growing $\sigma_{gas}$. Note that Boissier et al. found the same
behavior of $SFR_s$ for M~31 (see Figure~6 in  ~\cite{Bois06}).

It is essential that in the case when the surface gas density is
replaced by the volume density in the disc plane, the scatter of
curves in the diagram for individual galaxies reduces and they
become considerably better expressed (Figure~\ref{fig9}b).
Position of our Galaxy on the diagram is not distinguish
significantly from the other galaxies if not to take into account
that in the inner part where gaseous density is higher the star
formation rate remains rather moderate. This leads to the
non-monotonous character of curve. Boissier et al.~\cite{Bois03}
 also noted that the Schmidt law does not fulfill
in the inner part of the Galaxy.

\begin{figure}[h]
\epsfig{file=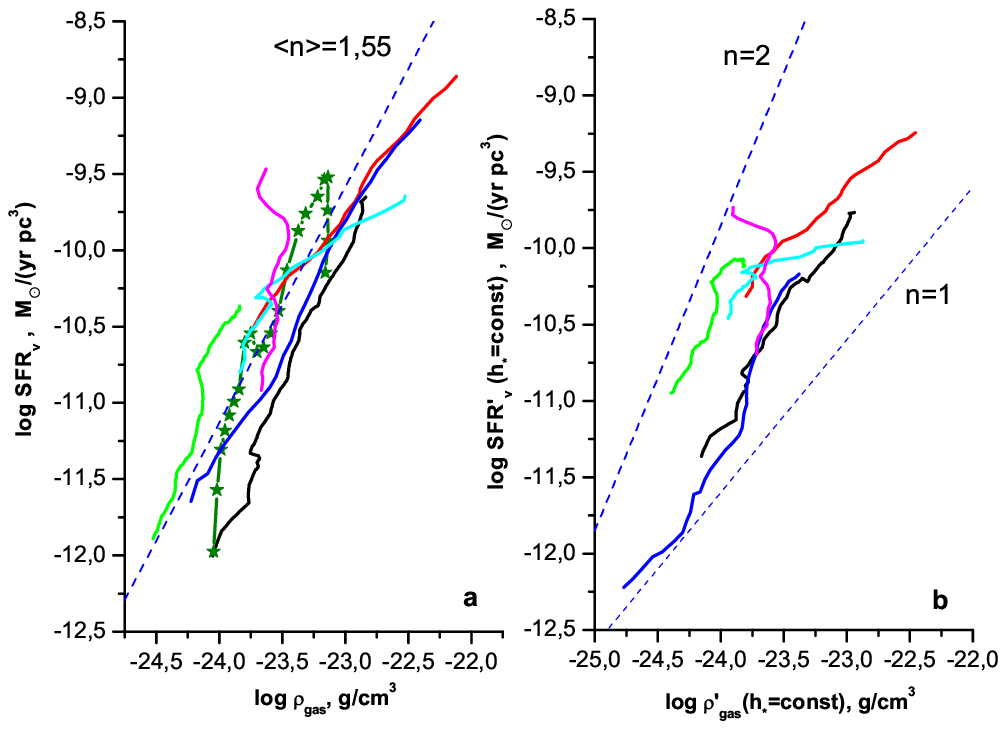,height=12cm,width=15cm} \caption{Star
formation rate over the unit of volume $SFR_v$ against the volume
gas density (classical Schmidt law). Left --- model with changing
stellar disc thickness, right --- model where stellar disc
thickness is fixed. Designations are the same as on the Figure~8.
In the panel (b) the Galaxy is not shown.} \label{fig10}
\end{figure}

If we want to consider the Schmidt law in its classical form
$SFR_v\sim \rho_{gas}^{n}$, the star formation rate must also be
related to the unit of volume.  Since star formation is tightly
connected with molecular gas, we assume $SFR_v=SFR/2h_{\rm H2}$ as
the mean volume density in regions of star formation, where
$h_{\rm H2}$ is the half thickness of molecular gas layer. The
star formation rate per unit volume $SFR_v$ is compared with the
total gas density in the disc plane in Figure~\ref{fig10}. The
diagram for marginally stable stellar discs whose thickness
changes with $R$ (case $a$), and the diagram for discs with a
fixed thickness (case $b$) are shown separately. In the latter
case, the curve for the Galaxy is not shown because we do not need
to make the assumption of the disc thickness constancy in its
case.
As it can be seen from Figure \ref{fig10}a, transition from
surface to volume gas and $SFR$ densitied decreases the difference
between the galaxies. For the constant thickness disc model
(Figure \ref{fig10}b) the dependencies for different galaxies are
not in too good concordance. However, even in this case, as in the
case of changing disc thickness,  all galaxies exept M~51 and
M~101 indicate the Schmidt law parameter $n\,>\,$1 (see
Table\,\ref{tab4}).
\begin{table}[h]
\caption{Parameter $n$ in the Schmidt law for the galaxies
considered. \emph{Columns}: (1) --- galaxies, (2), (3) and (4) ---
tangent of inclination angles of approximation lines in the
diagrams, their mean values and dispersions (in the
bottom).}\label{tab4}
\begin{center}
\begin{tabular}{c|c|c|c}
&&&\\
Galaxy&$n$ in $SFR_s\sim \rho_{gas}^{n}$&$n$ in $SFR_v\sim
\rho_{gas}^n$&$n$ in $SFR_v\sim \rho_{\rm{H_2}}^n$\\
&&&\\
\hline
&&&\\
(1)&(2)\,&(3)\,&(4)\,\\
&&&\\
\hline
&&\\
M~33&1,30&1,93&2,23\\
&&&\\
M~51&0,59&1,04&0,72\\
&&&\\
M~81&1,41&2,40&0,68\\
&&&\\
M~100&0,94&1,46&1,26\\
&&&\\
M~101&0,34&0,80&0,36\\
&&&\\
M~106&1,17&1,22&0,64\\
&&&\\
Galaxy&1,37&1,97&1,08\\
&&&\\
\hline
&&&\\
&$<$n$>$=1,02&$<$n$>$=1,55&$<$n$>$=0,99\\
&&&\\
&D=0,15&D=0,29&D=0,33\\
&&&\\
\hline
\end{tabular}
\end{center}
\end{table}

The gas volume density in the inner parts of M~51 and M~100 is the
highest and, as it is expected, the most high star formation rate
per  volume unit is observed there (Figure\ref{fig10}a). In spite
of general similarity between the curves for individual galaxies,
the range of $SFR_v$ remains rather high: a typical deviation from
the  mean linear dependence (dashed line) corresponds to a factor
of about 3. However, it is worth to mention that accuracies of
estimation of the gas density and star formation rate are of the
same magnitude.

Note that M~33 whose dependence $SFR_s(\rho_{gas})$ is especially
steep (parameter $N$ in Kennicutt\,--\,Schmidt law exceeds 3) does
not outlies considerably from the other galaxies in
Figure~\ref{fig10}. The high value of $N$ for this galaxy may be
naturally explained as the result of a strong flaring of gas layer
along the radius.

\begin{figure}[h]
\epsfig{file=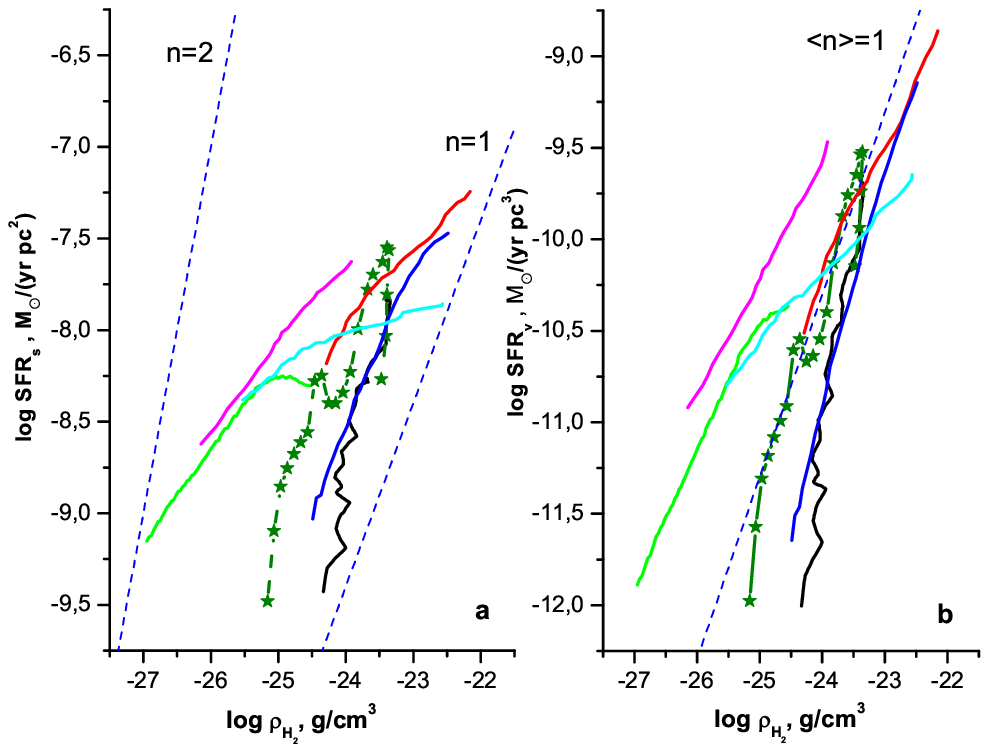,height=12cm,width=15cm} \caption{``Star
formation rate over molecular gas volume density'' diagram. Left
--- SFR is related to the unit of surface area, right --- to the unit of volume.
Designations are the same as in Figure~8.} \label{fig11}
\end{figure}

As in Figure~\ref{fig9}, M~106 does not follow the general
dependency: its gas density very slowly change along the radius,
whereas star formation rate decreases apart from the centre.
However, the situation for this galaxy becomes different if the
molecular gas density $\rho_{\rm H_2}$ (Figure~\ref{fig11}\,a,\,b)
is considered rather than the total gas density.
As one can see,  for all galaxies including M~106 star formation
rates increases with the increasing of the molecular gas density
${\rm H_2}$. In this case, the correlation also becomes more tight
if we consider  $SFR$ per unit volume rather than per unit area.
Note however that for the whole complex of galaxies $SFR_v$ better
correlates with total gas density, than with molecular gas density
(Figure~\ref{fig10}).

In Table~\ref{tab4}, the coefficients of inclination are presented
for the straight lines shown in Figures~\ref{fig9}b, \ref{fig10}a
and \ref{fig11}b and obtained by the least square method for each
galaxy (i.e. parameters $n$ in the Schmidt law)
\begin{equation}\label{1}
\lg SFR = n\,\lg\rho+const\,.
\end{equation}
Accuracy of the data for individual galaxies is different, so
there is no sence to mix up all the data for all galaxies to
calculate a common correlation coefficient or the mean value of
$n$. Instead, we calculate the values of $n$ for each galaxy
separately, and then find their mean value. The r.m.s. of $n$ that
illustrates the scatter of these coefficients is calculated in a
simple way:
\begin{equation}\label{587}
{\rm D}=\sum_{i=1}^{7}\frac{({\rm n}_i-<{\rm n}>)^2}{7}~.
\end{equation}
Values of n$_i$, $<n>$, and $D$ for the relations considered above
are given in Table~\ref{tab4}.

* Doshel do siuda, bol'she ne mogu. *

\section{Discussion and conclusions}

\hspace{0.6cm}As it follows from this paper, the assumption of
marginally stable stellar discs leads to conclusion that their
thickness, at least in some galaxies, changes significantly
(usually increases) along the  radius. The thickness of equlibrium
gas layer increases in all cases
--- either nearly linearly within a wide range of $R$ (M~33,
M~51, M~81 and M~101), or nonlinearly, with a positive second
derivative (M~100 and the Galaxy). Volume gas density in the
midplane in all cases decreases within the considered range of $R$
down to several units of $10^{-25}$ g/cm$^3$ . In all cases the
stellar density decreases steeper, than the gas density, so at the
peripheries of galaxies the gas midplane density becomes
comparable with the stellar density or even exceeds it.

To compare the half thicknesses (HWHM) with the densities of
stellar discs and gaseous layers in the galaxies it is convenient
to consider their parameters at a fixed radius, say, $R\approx
2R_0$ (see Table~\ref{tab5}).
\begin{table}[h]
\caption{Stellar disc and gas layers thicknesses and the densities
of gas at $R\approx 2R_0$. \emph{Columns}: (1) --- a galaxy, (2)
--- stellar disc thickness to radial scalelength ratio, (3) ---
atomic gas disc thickness, (4) --- molecular gas disc thickness,
(5) --- volume gas density (including helium).}\label{tab5}
\begin{center}
\begin{tabular}{ccccc}
&&&&\\
Galaxy&$h_*(2\,R_0)/R_0$&$h_{\rm HI}(2\,R_0)$\,,\,pc&
$h_{\rm H2}(2\,R_0)$\,,\,pc&$\rho_{gas}(2\,R_0)$\,,\,g/cm$^3$\\
&&&&\\
\hline
&&&&\\
(1)&(2)\,&(3)\,&(4)\,&(5)\,\\
&&&&\\
\hline
&&&&\\
M~33&0,18&87,8&57,6&$6,3\cdot10^{-24}$\\
&&&&\\
M~51&0,23&123,6&81,3&$3,5\cdot10^{-24}$\\
&&&&\\
M~81&0,34&129,2&85,7&$1,1\cdot10^{-24}$\\
&&&&\\
M~100&0,20&108,8&71,0&$6,1\cdot10^{-24}$\\
&&&&\\
M~101&0,19&173,8&114,8&$1,5\cdot10^{-24}$\\
&&&&\\
M~106$^*$&0,27&150,0&99,2&$2,2\cdot10^{-24}$\\
&&&&\\
Galaxy&0,11&100,3&61,4&$4,4\cdot10^{-24}$\\
&&&&\\
\hline
\end{tabular}
\end{center}
$^*$For M~106 all the values are given for R\,=\,1,5\,$R_0$.
\end{table}
As it follows from the Table, the relative thickness of stellar
disc $h_*/R_0$ at $R=2R_0$ lies within the range of 0.1\,--\,0.3.
Our Galaxy is the most thin ($h_*/R_0 \approx 0.1$) in our sample,
whereas the stellar discs of two giant early type spiral galaxies
(M~81 and M~106) are almost three times thicker. The thicknesses
of gas layers differ less than stellar ones in the galaxies. The
most thin gas layer (in M~33) is less than twice thinner than the
most thick one (in M~101). The gas density (at distance $2R_0$) in
all cases is equal to several units of 10$^{-24}$\,g/cm$^{-3}$
with no obvious dependence on the morphological type: in M~101
(Sc) it is approximately the same as in M~81 (Sab), and in our
Galaxy (Sb\,--\,Sbc) it is close to that in M~33.

The comparison of the gaseous density with the star formation
rates (figures\,\ref{fig9}a,~\ref{fig9}b, and~\ref{fig10}a) gives
an evidence that the replacement of surface gas density
(Figure~\ref{fig9}b) by the gas volume density leads to more tight
dependencies both for the ``surface'' and ``volume'' star
formation rates.  The reason for it is rather evident: star
formation takes place in a narrow layer near the midplane, hence
this process is sensitive namely to the volume gas density,
whereas the directly measured surface gas density depends, in
addition, to the gas layer thickness (more precisely --- to the
vertical density profile), which differs from galaxy to galaxy.

It is worth to remind that the regions that are very close to the
center (in all galaxies but M~33), or are located at the disc
peripheries (beyond the well defined spiral arms) need a special
study and were not considered in this paper. Indeed, here we
ignored the influence of the bulge on the vertical density
profiles. The accepted value of the turbulent velocity may also be
not suitable for all radii. A formal inclusion of the most inner
and outer regions into our diagrams leads to the resulting
dependencies ``gas density\,--\,$SFR$''\,\, which are less regular
and more different between galaxies.

Of the galaxies we consider, M~106 stands out by almost constant
density $\rho_{gas}$ between 4 and 10\,kpc. The increasing of
stellar disc thickness along the radius in this galaxy accompanies
by the increasing of the surface gas density out of the centre
within the wide range of $R$. By this reason this galaxy differs
from the others in the diagrams ``$SFR$\,--\,gas density\,''. The
similar abnormal behavior of this galaxy  on the Kennicutt-Schmidt
diagram was also revealed by Boissier et al.~\cite{Bois03}.
However the molecular gas density in M~106 behaves in the common
manner i.e. steeply decreases with $R$, that explains its
``normal'' position at the diagram $SFR_v\,-\,\rho_{\rm H2}$.
Note, that this galaxy is not typical in other aspects: it differs
from the others by the presence of the active nucleus, as well as
by the very active star formation in the central part of the disc.

The values of $n$ in the Schmidt law $SFR_v\,\sim\rho_{gas}^n$ for
the galaxies we consider are rather different (see
Table~\ref{tab4}). For all galaxies except M~101 the exponent
$n>1$, and its mean value is close to ``standard''\,\,value
$\approx 1.4$ for ``global''\,\,Kennicutt-Schmidt law (column (3)
of the Table). However, the mean value of $n$ approaches unit, if
to compare the $SFR_v$ with the molecular gas density (column (4)
in the Table).

 The most ``steep''\,\, relationships $SFR(\rho_{gas})$ are
obtained for M~33, M~81 and the Galaxy. In the case of
 M~33 it is caused by the steep decreasing of the $SFR$ with
the galactocentric distance (see Figure~\ref{fig8}). In the case
of M~81, it is the consequence of the atypically slow decrease of
the gas volume density along $R$ (see Figure~\ref{fig3}) when the
radial gradient of $SFR$ is moderate. For these three galaxies,
the exponent $n$ is close to the ``classical''\,\,value $n=2$,
suggested earlier by Schmidt.

In conclusion, parameter $n$ by no means can be considered as a
universal one. The  Schmidt law  has a very approximate nature,
and it fulfils much better when the  volume instead of the surface
gas density is used. In this case the exponent $n<2$ (with some
possible exceptions). It is important that in spite of a large
dispersion its value becomes closer to unit in the mean if to
replace the total gas density $\rho_{gas}$ with the molecular gas
density $\rho_{\rm H_2}$.

It is evident that relationship between the volume gas density and
$SFR$ for galaxies should not be too tight because the process of
star formation depends on a number of parameters besides the mean
gas density. The latter seems to be a crucial factor only for the
most dense gas such as the gas in the nuclei of molecular clouds
where the HCN radiation comes from. Indeed, as observations show,
the star formation rate depends linearly  on the gas mass
determined by the HCN line intensity, i.e for the most dense
molecular gas $n\approx 1$ (Gao and
Solomon,~\cite{gao04a,gao04b}). Although the number of galaxies we
considered is small, the results allow to propose that the value
of $n$ is on average close to unit even for the less dense
molecular gas, that reveals itself in CO line. Since $n>1$ for the
total ($HI+H_2$) gas density, one may conclude that not only star
formation rate, but also the efficiency of star formation ($SFR$
per gas mass unit) decreases along with $\rho_{gas}$. In other
words, the less is the gas density, the longer time the gas
remains in the rarified atomic state (that is a time scale of gas
consumption is larger). It agrees with general conception,
according to which the fraction of the interstellar gas which
takes active part in the process of a star formation, decreases
when the mean total density of gas becomes lower (see the
discussion in~\cite{WB02,elm03}).

The other important factor, besides the gas density, that
determines the star formation rate at a given radius $R$ is the
surface density of the old stellar disc $\sigma_{*}$ (see
papers~\cite{ZasAbr06,BdeJ00,DR94}). As it was shown by Zasov and
Abramova~\cite{ZasAbr06} on the example of four well-studied
galaxies, local star formation efficiency defined as
$SFE\,=\,SFR$/$\sigma_{gas}$  changes approximately as
$\sigma_{*}^{\footnotesize 0,7}$ in a wide interval of $R$.
Partially this relationship may be explained by higher volume gas
density (for a given surface density) in those regions, where
$\sigma_{*}$ is higher (that is closer to galaxy center).
Nevertheless this relationship can not be reduced to the simple
exponential Schmidt law (neither ``volume'' nor ``surface'' one)
because the clearly defined relationship between the gas density
and surface disc density is absent. It evidences the existence of
 more deep connection between the present-day star formation on the
one hand and already formed stellar disc and the gas density on
the other hand, which cannot be described by simple empiric laws.

\bigskip

The authors thank Igor V. Abramov for assistance in the numerical
solutions and D.Bizyaev for helpful discussion.

\bigskip

This work was supported by the Russian Fond of Basic Researches
grant 07-02-00792.


\begin{thebibliography}{60}

\bibitem{Schmidt59}
M.~Schmidt, ApJ, \textbf{129}, 243 (1959)

\bibitem{Madore74}
B.~F.~Madore, S.~van~den~Bergh, D.~H.~Rogstad, ApJ, \textbf{191},
317-322 (1974)

\bibitem{Kennicutt89}
R.~C.~J.~Kennicutt, ApJ, \textbf{344}, p. 685-703, (1989)

\bibitem{WB02}
T.~Wong, L.~Blitz, ApJ, \textbf{569}, 157 (2002)

\bibitem{Bois03}
S.~Boissier, N.~Prantzos, A.~Boselli, G.~Gavazzi, MNRAS,
\textbf{346}, 1215 (2003)

%\bibitem{\check{Bois}03}
%S.~Boissier, N.~Prantzos, A.~Boselli, G.~Gavazzi, MNRAS,
%\textbf{346}, 1215 (2003)

\bibitem{Schu06}
K.~F.~Schuster, C.~Kramer, M.~Hitschferd et al, A\&A,
\textbf{461}, 143 (2007)

\bibitem{tut06}
A.~V.~Tutukov, Astronomy Reports, \textbf{50}, 526 (2006)

\bibitem{HeyerAll04}
M.~H.~Heyer, E.~Corbelli, S.~E.~Schneider, J.~S.~Young, ApJ,
\textbf{602}, Issue 2, 723-729 (2004)

\bibitem{Kennicutt98}
R.~C.~Kennicutt, ApJ, \textbf{498}, 541 (1998)

\bibitem{li+06}
Y.~Li, M.-M.~Mac Low, R.~S.~Klessen, ApJ, \textbf{639}, 879 (2006)

\bibitem{elm02}
B.~G.~Elmegreen, ApJ, \textbf{577}, 206 (2002)

\bibitem{krumholz05}
M.~P.~Krumholz, C.~F.~McKee, ApJ, \textbf{630}, 250 (2005)

\bibitem{Gerristen97}
J.~P.~E.~Gerristen, V.~Icke, A\&A, \textbf{325}, 972 (1997)

\bibitem{dib06}
S.~Dib, E.~Bell, A.~Burkert, ApJ, \textbf{638}, 797 (2006)

\bibitem{Cox05}
D.~P.~Cox, ARA\&A, \textbf{43}, Issue 1, 337-385 (2005)

\bibitem{NJ02}
C.A. Narayan, C.J. Jog, A\&A, \textbf{394}, 89 (2002)

\bibitem{Leda}
HyperLeda: http://leda.univ-lyon1.fr/

\bibitem{BdeJ01}
E.~F.~Bell, R.~S.~de~Jong, ApJ, \textbf{550}, 212 (2001)

\bibitem{Bottema93}
R.~Bottema, A\&A, textbf{275}, No. 1 (1993)

\bibitem{ZasAll04}
A.V.Zasov, A.V.Khoperskov, and N.V.Tyurina, Astronomy Letters,
\textbf{30(9)}, 593-602 (2004)

\bibitem {HZT03}
A.V.Khoperskov, A.V.Zasov, and N.V.Tyurina, Astronomy
Reports,\textbf{47(5)}, 357-376 (2003)

\bibitem{vdKruit88}
P.~C.~Kruit, A\&A, \textbf{192}, no. 1-2, 117-127 (1988)

\bibitem{deGrPel99}
R.~de Grijs, R.F.~Peletier, MNRAS, \textbf{310}, Issue 1, 157-167
(1999)

\bibitem{celnik79}
W.Celnik, K.Rohlfs, E.Braunsfurth, A\&A, 76, 24 (1979)

\bibitem {LewFrem89}
Lewis J.R. \& Freeman K.C., AJ, \textbf{97}, 139 (1989)

\bibitem {Kh03}
A.V.Khoperskov and N.V.Tyurina, Astronomy Reports,\textbf{47(6)},
443-457 (2003)

\bibitem {main}
S. Boissier, A. Boselli, V. Buat, J. Donas, B. Milliard, A\&A,
\textbf{424}, 465 (2004)

\bibitem {HBI03}
H. Hirashita, V. Buat, A.K. Inoue, A\&A, \textbf{410}, 83 (2003)

\bibitem {BBGB02}
V. Buat, A. Boselli, G. Gavazzi, C. Bonfanti, A\&A, \textbf{383},
801 (2002)

\bibitem{Iglesias04}
J. Iglesias-Paramo, V. Buat, J. Donas, A. Boselli, B.Milliard,
A\&A, \textbf{419}, 109 (2004)

\bibitem {ZasAbr06}
A.V.Zasov and O.V.Abramova, Astronomy Reports, \textbf{50(11)},
874-886 (2006)

\bibitem {Bronfetal00}
Bronfman et al, A\&A, \textbf{358}, 521 (2000)

\bibitem{Rana91}
N.C.Rana, Ann. Rew. A\&A, \textbf{29}, 129, 1991

\bibitem{Bois06}
S.Boissier, A.Gil de Paz, A.Boselli et al, astro-ph 0609071 (2006)

\bibitem{gao04a}
Y. Gao, P.M. Solomon, ApJ, 606, 271 (2004a)

\bibitem{gao04b}
Y. Gao, P.M. Solomon, ApJS, 152, 63 (2004b)

\bibitem{elm03}
B.G.Elmegreen, ApSS, 284, 819 (2003)

\bibitem{BdeJ00}
E.F.Bell, R.S.deJong, MNRAS, \textbf{312}, 497 (2000)

\bibitem{DR94}
M.~A.~Dopita, S.~D.~Ryder, ApJ, \textbf{430}, 142 (1994)

\bibitem{Lauer}
T. Lauer, S. Faber, E. Ajhar, C. Grillmair, P. Scowen, AstronJ,
\textbf{116}, 2263 (1998)

\bibitem {RegVog}
M.W. Regan, S.N. Vogel, ApJ, \textbf{434}, 536 (1994)

\bibitem {Corb}
E. Corbelli, MNRAS, \textbf{342}, 199 (2003)

\bibitem {Baggett}
W.E. Baggett, S.M. Baggett, K.S.J. Anderson, ApJ, \textbf{116},
1626 (1998)

\bibitem {STHTTKT99}
Y. Sofue, T. Tutui, M. Honma, A. Tomita, T. Takamiya, J. Koda, Y.
Takeda, ApJ, \textbf{523}, 136 (1999)

\bibitem {Rots}
A.H. Rots, A\&A, \textbf{45}, 43 (1975)

\bibitem {Beckman}
J.E. Beckman, R.F. Peletier, J.H. Knapen, R.L.M. Corradi, L.J.
Gentet, ApJ, \textbf{467}, 175 (1996)

\bibitem {vanAlb80}
van Albada G.D., A\&A, \textbf{90}, 123 (1980)

\bibitem {de Jong}
R.S. de Jong, A\&ASuppl, \textbf{118}, 557 (1996)

\bibitem {Knapen}
J.H. Knapen, R.S. de Jong, S. Stedman, D.M. Bramich, MNRAS,
\textbf{344}, 527 (2003)

\bibitem {Flores}
R. Flores, J.R. Primack, G.R. Blumenthal, S. M. Faber, ApJ,
\textbf{412}, 443 (1993)

\bibitem {HeraudSimien}
Ph. H\'eraudeau\&F. Simien, A\&ASS, \textbf{118}, 111 (1996)

\bibitem {Sanchezetal}
M. S\'anchez-Portal, A.I. D\'iaz, R. Terlevich, E. Terlevich, M.
\'Alvarez \'Alvarez, I. Aretxaga, MNRAS, \textbf{312}, 2 (2000)

\end{thebibliography}
\end{document}